\documentclass[12pt]{iopart}
\usepackage[a4paper,top=2.5cm,bottom=2cm,left=2cm,right=2cm]{geometry}

\usepackage[symbol]{footmisc}
\newcommand\correspondingauthor{\footnote{Corresponding author: \texttt{v.gupta-2@tudelft.nl}}} 
\usepackage[colorlinks=true, allcolors=blue]{hyperref}
\usepackage{float}
\usepackage{graphicx}
\usepackage{subcaption}
\usepackage{enumitem}

\begin{document}

\title[Piezoelectric Shear Actuators for Vibration
Control in Sandwich Beams]{Harnessing Piezoelectric Shear Actuators for Vibration Control in Sandwich Beams}
\author{Mark Baken$^1$, Vivek Gupta\correspondingauthor$^1$, Bas Jansen$^2$ and S.Hassan HosseinNia$^1$}

\address{$^1$ Department of Precision and Microsystems Engineering,
Delft University of Technology, Delft, The Netherlands}
\address{$^2$ ASML, Veldhoven, The Netherlands}

%\ead{v.gupta-2@tudelft.nl}
% \author{Content \& Services Team}

% \address{IOP Publishing, No. 2 The Distillery, Avon Street, Bristol BS2 0GR, UK}
% \ead{customerservices@ioppublishing.org}
\vspace{10pt}
\begin{indented}
\item[]March, 2025
\end{indented}

\begin{abstract}
Our study found that integrating shear piezo-transducers inside the beam offers a compact and efficient solution that enables localized damping control without compromising structural integrity. However, the conventional approach of placing the piezos outside the substrate faces challenges and limited accessibility to industrial applications. We determine damping performance for long and slender sandwich beam structures utilizing active vibration control by internally placed piezoelectric shear sensors and actuators. Experimental and numerical results are presented for a clamped-free sandwich beam structure constructed with two stainless steel facings composed of a core layer of foam and a piezoelectric shear-actuator and sensor. This approach of internal actuator and sensor tends to tackle the problems within (high-tech) systems, i.e. mechanical vibrations, a limited amount of design volume, and vulnerability of externally placed piezoelectric transducers to outside conditions. By this new internal sensor-actuator approach, this study addresses a significant gap in the literature. 
The location of the sensor and actuator has been defined by numerical investigation of the \textit{modal shear strain} and the \textit{effective electro-mechanical coupling coefficient}. The frequency response of the sandwich beam structure has been evaluated using both numerical and experimental investigation. Positive Position Feedback has been employed on the numerical response to simulate the damping performance for the fundamental mode. Different controller gains have been used to analyze the trade-off between effective resonance suppression and increased low-frequency gain. The tip vibrations at the fundamental mode have been reduced from 5.01 mm to 0.34 mm amplitude at steady state, which represents a significant reduction. 

\end{abstract}

% Include a list of keywords after the abstract 

%
% Uncomment for keywords
\vspace{2pc}
\noindent{\it Keywords}: Active vibration control, Piezoelectric shear actuator, Piezoelectric shear sensor, Sandwich beam, Damping, PPF. 
%
% Uncomment for Submitted to journal title message
%\submitto{\JPA}
%
% Uncomment if a separate title page is required
%\maketitle
% 
% For two-column output uncomment the next line and choose [10pt] rather than [12pt] in the \documentclass declaration
%\ioptwocol
%
\section{INTRODUCTION}
\label{sec:intro}  % \label{} allows reference to this section

The high-tech industry has an ever-increasing demand for high throughput and positioning accuracy of semi-conductor machines, which is mainly driven by the ability to produce microelectronic chips that are more powerful, yet smaller and at the same time even low-priced \cite{zhang2014silicon}. The high throughput results in machines with critical components that exhibit fast motion and high accelerations, which makes these systems, and other subsystems, of precision machines more sensitive to mechanical vibrations \cite{butler2011position}. Examples of such critical components are \textit{wafer handlers} \cite{ajjaj2022vibration} and \textit{reticle masking units}. These systems are characterized by their small thickness and large length (long slender structures). Basic Active Vibration Control (AVC) strategies, such as Integral Force Feedback (IFF) \cite{fleming2010integrated}, Direct Velocity Feedback (DVF) \cite{balas1979direct} and Positive Position Feedback (PPF) \cite{fanson1990positive}, have proven to be a powerful, yet simple solution for suppressing these mechanical vibrations. PPF control is often combined with collocated piezoelectric actuators and sensors, especially for suppressing resonant modes of long and slender structures \cite{moheimani2006piezoelectric} and as shown in recent studies of placing transducers in periodic fashion \cite{kaczmarek2024active, kaczmarek2024bandgap}. The collocated configuration ensures pole-zero interlacing, making the plant inherently stable \cite{preumont_vibration_2011}. The wide utilization of piezoelectric transducers (piezo's) can be attributed to their numerous advantages \cite{aridogan2015review}. Due to their high power density, piezoelectric transducers enable a compact system design and therefore have relatively small impact on a system's mass, mode shape and resonant frequency. 
\\ 
Many of the preceding techniques are based essentially on the surface bonded piezoelectric extension transducer, often referred to as extension mechanism. For long slender beam structures, the dynamics of the bonded extension transducer can directly be related to the modal bending strain by the Euler-Bernoulli beam theory \cite{preumont_vibration_2011}. Combining this with collocated placement, the optimal location of the actuator and sensor are found at the top- and bottom-surface of beam structures \cite{gupta2010optimization}.  
\\
However, the design volume of precision systems is often limited and therefore placement of the actuator and sensor on the outer surface of the structure is not always possible. In addition, externally placed components are vulnerable to outside conditions, which may effect the durability of delicate components such as piezoelectric transducers or the  the piezo and glue used bond methods may impact the (vaccuum) environment. A perfect solution to address those two problems is by placing the actuator and sensor internally, on the inside of the solid beam structure. By making use of this internal approach the design volume of the beam structure will be used optimally and the most delicate components, the piezoelectric actuators and sensors, are being protected by the structure itself.  
\\
Internal placement of the actuator and sensor requires a different type of piezoelectric transducer. This can be derived from the fact that the modal bending strain will decrease towards zero towards the neutral axis of the clamped-free beam structure \cite{gupta2010optimization}. At the neutral axis the piezoelectric extension sensor will measure zero modal bending strain and the extension actuator cannot counter-act nor measure the modal bending strain effectively. This makes the active vibration control loop between sensor and actuator ineffective.
\\
Multiple studies have investigated this problem by placing piezoelectric shear transducers instead of extension extension transducers inside the core of the beam structure. 
Benjeddou et. al.\cite{benjeddou1999new} proposed a finite element model with extension and shear-induced actuation. The static analysis and free vibration analyses of sandwich beams were carried out to see the behavior of both extension and shear. It was shown that the shear actuator generates tip displacement by inducing transverse shear strains between two stiff outer layers. The layers were connected by the piezo shear actuator and a foam core-layer. 
Trindade et. al. \cite{trindade1999parametric} subsequently carried out a vibration control study on sandwich beams with extension and shear actuators. A linear quadratic regulator (LQR) based optimal control was used with state feedback to find the effect of shear actuators in controlling the bending vibration of sandwich beams. It was stated that the sandwich structure should be constructed with a relatively soft core-layer to achieve good damping performance. This could be explained by the fact that the shearing between the stiff outer layers increases and distributes more equally over the length of the structure for core-layers with a low stiffness. In this way, the shear actuator can counter act the modal shear strains more effectively. 
Different finite element methods have investigated the damping performance of the shear actuator for different boundary conditions. \cite{raja_active_2002} \cite{benjeddou2001piezoelectric} \cite{benjeddou2002two}.
Subsequently, Baillargeon et. al.\cite{vel_analysis_2005} experimentally investigated the damping performance of the shear actuator by basic AVC approaches. Despite, a surface
bonded strain sensor was used instead of a piezoelectric shear sensor to close the feedback control loop. This leaves a research gap in both numerical and experimental investigation of basic AVC utilizing piezoelectric shear actuators and piezoelectric shear sensors.  
\\
The aim of this research is to demonstrate the damping performance for long slender sandwich beam structures by internally placed piezoelectric shear actuators and sensors, utilizing basic active vibration feedback control. This is done by both numerical and experimental investigation.
\\
The outline of this paper is defined as followed. First, the location of the shear sensor and shear actuator has been defined by numerical investigation of the modal shear strain and the effective electro-mechanical coupling coefficient. Second, the frequency response of the
sandwich beam structure has been evaluated by both numerical and experimental investigation. Third, Positive Position Feedback is implemented on the numerical response of the finite element model to demonstrate the damping performance on the fundamental mode.

\section{Methods and Materials}
This section gives a description of the methods used in this paper, which includes three main parts. (1) Description of the \textit{modal shear strain} and \textit{Effective Electro-Mechanical Coupling Coefficient} that are used to find the location of the piezoelectric shear sensor and piezoelectric shear actuator by numerical simulations. (2) Description of the experimental work. In the experimental work open-loop identification of a sandwich beam structure utilizing both a piezoelectric shear sensor and shear actuator have been performed with the validations. These results were used to verify the numerical results of the finite element method (FEM) model described in (3). (3) Description of the numerical model used to simulate the active vibration control performance of a sandwich beam structure utilizing both a piezoelectric shear sensor and piezoelectric shear actuator. The FEM models utilize the commercial analysis software of \textit{COMSOL Multiphysics 6.2}.

\subsection{Constitutive Equations}
The constitutive equations of the shear piezoelectric transducer have been used to relate the electro-mechanical behavior of the piezoelectric transducer to the sandwich beam structure, see Eq. \ref{eq:const_eq}.

\begin{equation} \label{eq:const_eq}
	\left[
	\begin{array}{c}
		\mathrm{T} \\
		\mathrm{D}	 
	\end{array}
	\right]
	=
	\left[
	\begin{array}{cc}
		\frac{1}{s^E} & -\frac{d}{s^E} \\
		\frac{d}{s^E} & \epsilon^T - \frac{d^2}{s^E} 
	\end{array}
	\right]
	\left[
	\begin{array}{c}
		\mathrm{S} \\
		\mathrm{E}
	\end{array}
	\right]
\end{equation}

\begin{figure} [h!]
	\centering
	\begin{tabular}{c}
		\includegraphics[height=1.8cm]{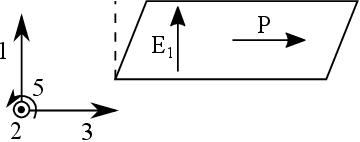}
	\end{tabular}
	\caption[example] 
	%>>>> use \label inside caption to get Fig. number with \ref{}
	{ \label{fig:shear_piezo} 
		Graphical representation of a sheared piezoelectric shear transducer. Poling-direction $P$ is defined in 3-direction. Direction 1, 2, and 3 refer to translations. Direction 5 refers to rotation about the 2-axis. In the proceeding of this paper the 1, 2 and 3-directions of the piezoelectric transducers are parallel to the $z$, $y$ and $x$direction respectively.}
\end{figure}

% \begin{equation} \label{eq:const_eq}
% 	\begin{bmatrix}
% 		T \\
% 		D	 
% 	\end{bmatrix}
% 	=
% 	\begin{bmatrix}
% 		\frac{1}{s^E} & -\frac{d}{s^E} \\
% 		\frac{d}{s^E} & \epsilon^T - \frac{d^2}{s^E} 
% 	\end{bmatrix}
% 	\begin{bmatrix}
% 		S \\
% 		E
% 	\end{bmatrix}
% \end{equation}
\noindent
Where $S \; [m/m]$ is the \textit{mechanical strain}, $E \; [V/m]$ the \textit{electrical field}, $T \; [N/m^2]$ the \textit{mechanical stress} and $D \; [C/m^2]$ the \textit{electric charge displacement}. The material constants $s^E \; [m^2/N]$, $d \; [m/V] \mid [C/N]$ and $\epsilon^T \; [F/m]$ denote the \textit{mechanical compliance}, \textit{piezoelectric charge} and \textit{permitivity} of the piezoelectric transducer.   
By this definition, the applied \textit{electrical field} $E$ on the piezoelectric actuator results in (shear) \textit{stress} $T$ exerted on the beam structure, which is directly related to the exerted force. For the piezoelectric sensor, the (shear) \textit{strain} $S$ generated by the beam structure is related to \textit{electric charge displacement} $D$. The poling direction of the piezoelectric shear transducer is defined in the 3-direction, see Fig.  \ref{fig:shear_piezo}. In the proceeding of this paper the 1, 2 and 3-directions of the piezoelectric transducers are parallel to the $z$, $y$ and $x$ direction respectively. The piezoelectric material constants such as compliance ${s_{55}}^E$, piezo coefficient $d_{15}$, permitivity $\epsilon^T$ and density $\rho_p$ have been derived experimentally according to the method of Sherrit \cite{sherrit2007characterization}, see appendix \ref{Appendix:A_piezo}.    
\subsection{Optimal location piezoelectric transducer}
\label{sec:optimal_loc}
\begin{figure} [H]
	\centering
	\begin{tabular}{c}
		\includegraphics[height=4.5cm]{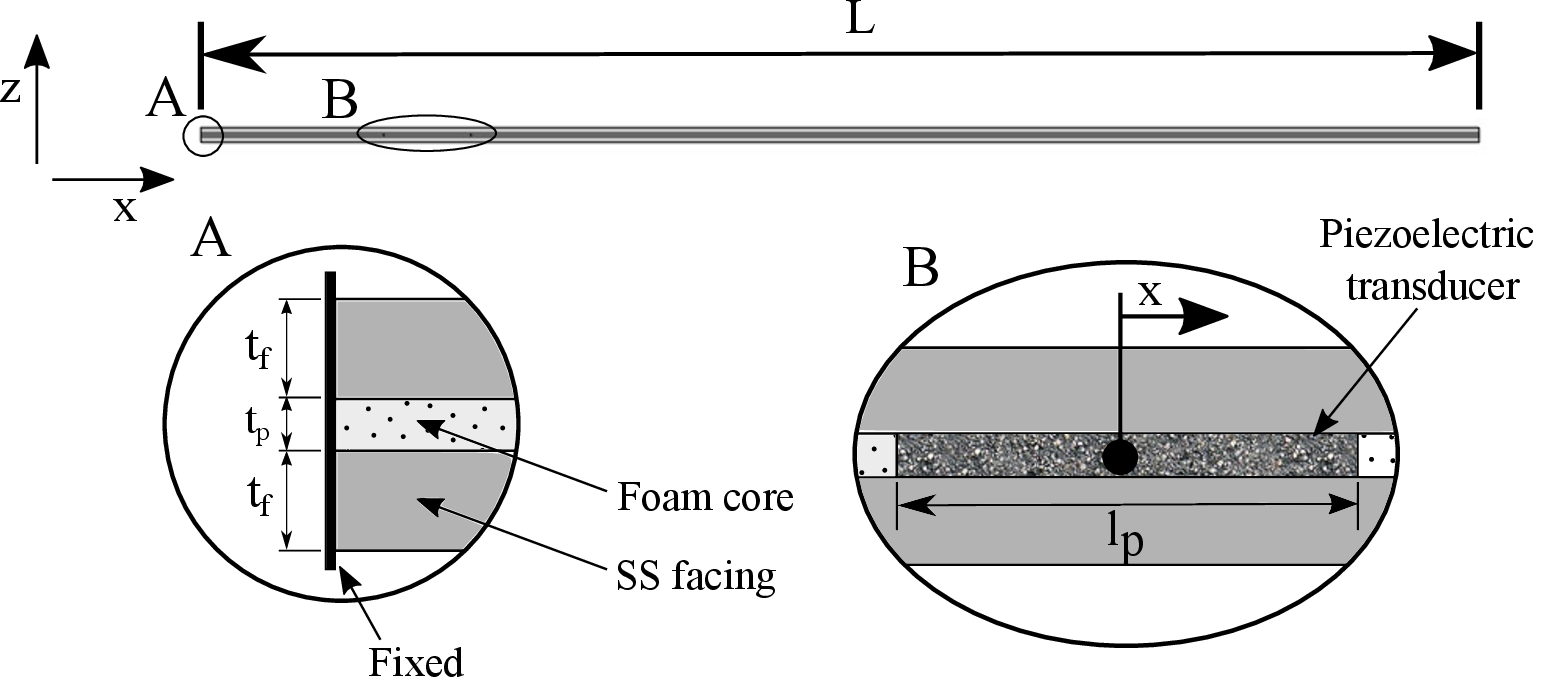}
	\end{tabular}
	\caption[example] 
	%>>>> use \label inside caption to get Fig. number with \ref{}
	{ \label{fig:model_optx} 
		Graphical representation of the structure utilized for the \textit{modal shear strain} and \textit{Effective Electro-Mechanical Coupling Coefficient (EMCC)} analysis. In zoom-section A the clamped base of the structure is shown. Zoom-section B presents the piezoelectric shear transducer, where $x$ is changed parametrically in the \textit{EMCC} analysis. Parameter values and material constants can be found in appendix \ref{Appendix:B_parameter}.}
\end{figure}

Two different methods have been used in this paper to find the optimal x-location of the piezoelectric transducer. The x-location refers to the location of the piezo in the length direction of the beam structure. This has been done according to the modal shear strain of the core-layer and the Effective-EMCC of the internally placed piezo. The next two sections will explain these methods in further detail. A graphical representation of the finite element model utilized is shown in Fig. \ref{fig:model_optx}. 
\subsubsection{Modal shear strain}
\begin{figure} [H]
	\centering
	\begin{tabular}{c}
		\includegraphics[height=4.5cm]{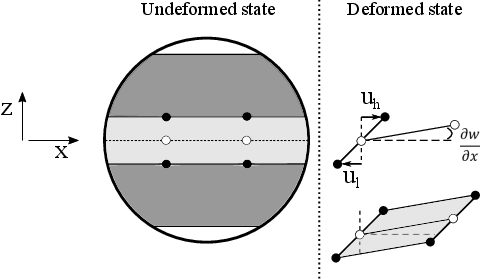}
	\end{tabular}
	\caption[example] 
	%>>>> use \label inside caption to get Fig. number with \ref{}
	{ \label{fig:modal_strain} 
		Graphical representation of the shear strain inside the core-layer of a sandwich beam structure. The undeformed state shows a zoomed section of a sandwich beam structure, where the dots represent the nodes of the FEM model. The deformed state shows an arbitrary displacement field of the core-layer, where $U_h$, $U_l$ and $\frac{\partial w}{\partial x}$ are used to calculate the shear strain.}
\end{figure}

A method to find the optimal $x-$location of the piezoelectric transducer is by evaluating the modal strains of the host structure, which is extensively used for extension mechanisms \cite{gupta2010optimization}. For these systems the bending strain is directly related to the extension and contraction of the piezo element \cite{preumont_vibration_2011}. However, embedded shear piezo's induce or measure transverse shear strains instead of bending strains directly and therefore the (modal) shear strain of the core-layer is related to the shearing of the piezo element \cite{benjeddou1999new}. In this paper, a method has been developed to find the optimal location according to the modal shear strain of the core-layer. A graphical representation of the structure is shown in Fig. \ref{fig:model_optx}. Note that the piezoelectric transducer is excluded from the structure in this analysis. The simulation is based on the eigenvalue problem as shown in Eq. \ref{eq:eigenvalue}.
\begin{equation} \label{eq:eigenvalue}
	(\mathbf{K_s}-\omega^2\mathbf{M})\mathbf{u} = 0
\end{equation}
Where $\mathbf{K_s}$ is the stiffness matrix, $\mathbf{M}$ the mass matrix, $\mathbf{u}$ the eigenmode displacement vector and $\omega$ the angular frequency. Within the analysis, $\mathbf{u}$ has been orthogonalized with respect to the mass matrix $\mathbf{M}$.
\\
\\
To be able to consider the modal shear strain ($S_i(x)$ for mode-$i$) in the core-layer of the structure only, the equation for the modal shear strain is defined in Eq. \ref{eq:modal_strain}. The optimal $x-$location of mode-$i$ can be found by finding the maxima of $ \mid S_i(x)\mid$.
\begin{equation} \label{eq:modal_strain}
	S_i(x) = \frac{U_d}{\frac{1}{2} h} + \frac{\partial W}{\partial x}, \quad \mathrm{where} \quad U_d = \frac{U_h - U_l}{2}
\end{equation}
Where $U_h$ and $U_l$ represent the displacement of the nodal points in $x-$direction, $w$ the displacement in $z-$direction and $h$ the thickness of the core-layer, as shown in Fig. \ref{fig:modal_strain}. The first part of Eq. \ref{eq:modal_strain} defines the vertical rotation and the second part the horizontal rotation of the core layer, these two combined forms the shear strain. This approach is inspired on the shear strain definition used for other FEM sandwich beam structures \cite{benjeddou1999new} that is defined by the \textit{Timoshenko method}. \cite{bauchau_structural_2009} \cite{ochsner2021classical} An alternative method for defining the $x-$location is described in the next subsection. 

\subsubsection{Effective Electro-Mechanical Coupling Coefficient}
A second method to find the optimal $x-$location of embedded shear piezo's is according to the Effective Electro-Mechanical Coupling Coefficient (Effective-EMCC and \autoref{eq:EMCC}) \cite{article} is based on the open and short circuit eigenfrequency of the piezoelectric transducer that is embedded inside/on the host structure \cite{holterman2012introduction}. Since the piezo is coupled to the structure, the modal Effective-EMCC describes the efficiency for transforming strain energy from the piezo patch into modal vibration energy of the host structure. An high modal Effective-EMCC means that the piezo is well coupled to the structure in order to excite/observe the modes of the structure.
\\
For this analysis the piezoelectric shear transducer is located inside the core-layer of the sandwich structure, as shown in Fig. \ref{fig:model_optx}. A parametric evaluation for the Effective-EMCC over the $x-$location was performed, where the length, thickness and width of the piezo were set fixed. The equation used to calculate the Effective-EMCC of mode-$i$ of the structure is shown in Eq. \ref{eq:EMCC}, where the optimal $x-$location can be found by finding the maxima of $K_i^2(x)$. 
\begin{equation} \label{eq:EMCC}
	K_i^2(x)=\frac{{{\omega}_{oc,i}}^2 - {{\omega}_{sc,i}}^2}{{{\omega}_{oc,i}}^2}
\end{equation}
\noindent
The open-circuit frequency $\omega_{oc,i}$ and short-circuit frequency $\omega_{sc,i}$ were derived from the impedance of the piezoelectric transducer \cite{holterman2012introduction}. The impedance was found by evaluating its harmonic response over a defined range of frequencies.

\subsection{Experimental validations}
In this section the experimental setup utilized in this paper is described, see Fig. \ref{fig:setup}. This setup has been used to identify the frequency response of a sandwich beam structure consisting of both a piezoelectric shear sensor and piezoelectric shear actuator. Results gained from the open-loop experiments have been used to validate the FEM model employed for active vibration control.

\subsubsection{Sandwich beam design}
\begin{figure}[h!]
	\centering
	\begin{minipage}{0.6\textwidth}
		\centering
		\includegraphics[height=2cm]{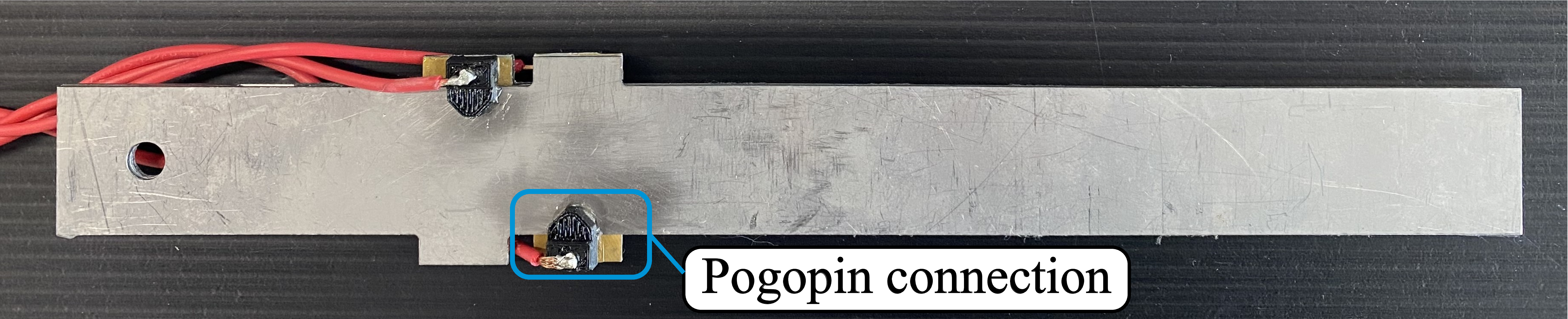} 
		\subcaption{}
		\label{fig:sandwich_a}
	\end{minipage}
	\vspace{0.5cm} % Adds vertical space between the two figures
	\begin{minipage}{0.7\textwidth}
		\centering
		\includegraphics[height=2.5cm]{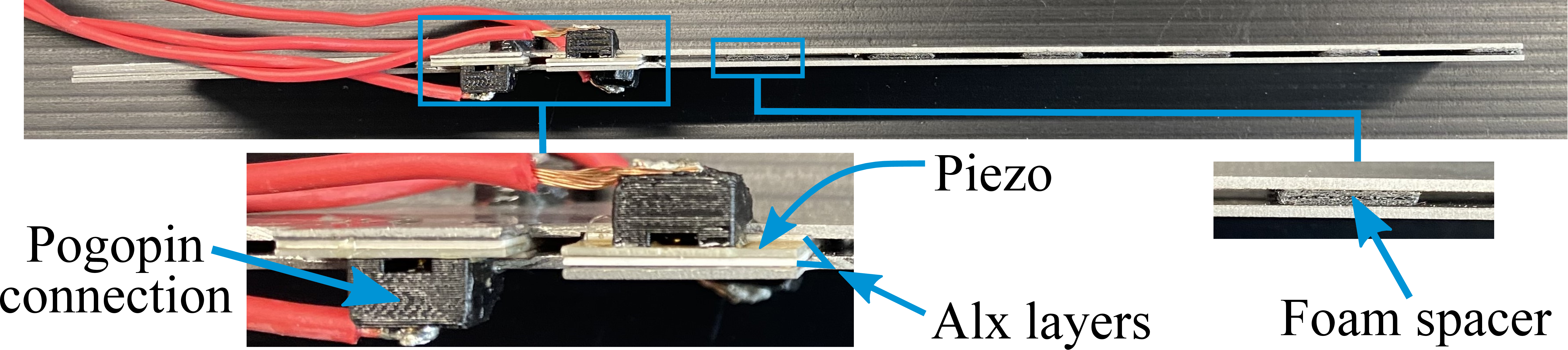}  
		\subcaption{}
		\label{fig:sandwich_b}
	\end{minipage}
	\caption{Sandwich beam structure employed in the experimental work. The top and side view of the sandwich beam are shown in Fig. \ref{fig:sandwich_a} and \ref{fig:sandwich_b}, respectively. (a) Top view of the structure. The electrical wires of the piezo are connected by pogo-pin connections, present in the black 3D-printed housing. (b) Side view of the structure. Piezo elements are electrically isolated from the facings by alumina (alx) layers. Foam spacers are present in the core layer of the structure.}
	\label{fig:sandwich} % Moved label outside of the caption
\end{figure}

In Fig. \ref{fig:sandwich} the sandwich beam structure employed in the experimental tests is shown. A graphical representation, containing all parameters, can be found in Fig. \ref{fig:FEM_model2}.  
\noindent
To prevent the piezo's from being short-circuited by the stainless steel (SS) facings, alumina layers on both the top- and bottom area were added to the piezo, see Fig. \ref{fig:sandwich_a}. The electrical wires are connected to the piezoelectric transducers by pogo-pin connections that are mounted inside the black 3D-printed housing, see Fig. \ref{fig:sandwich_a} and \ref{fig:sandwich_b}. This press contact-connection was preferred over soldering since delamination had been observed between the electrode layer of the piezo and the piezo core material in previous designs. A small air-gap, denoted by distance $r$ in Fig. \ref{fig:FEM_model2}, is maintained to ensure electrical isolation between the actuator and sensor. In addition, air-gaps are present in the foam core layer, which were thought to decrease the structural damping. The foam layer presents industrial double-sided polyethylene \textit{Tesa® 62932} foam tape. The foam tape required the low-stiffness property which is essential for the working principle of the shear mechanism \cite{trindade1999parametric}. However, due to the lack of exact material data, the material specifications of the foam layer have been estimated within the FEM simulation. Spacers are present in the core-layer, near the clamping region shown in zoom-section A in Fig. \ref{fig:FEM_model2}. These spacers prevent the facings from being squeezed together when the sandwich beam was mounted to the setup. The SS spacers and facings around zoom-section A, as well as the SS facing, alumina isolation and piezoelectric transducers around zoom-section B, are connected by \textit{EPO-TEK® 301-2} epoxy glue. Parameter values of the sandwich beam and beam material constants can be found in appendix \ref{Appendix:B_parameter}.

\subsubsection{Experimental setup}
\begin{figure}[h!]
	\centering
	\begin{minipage}[h]{0.49\textwidth}
		\centering
		\includegraphics[height=7cm]{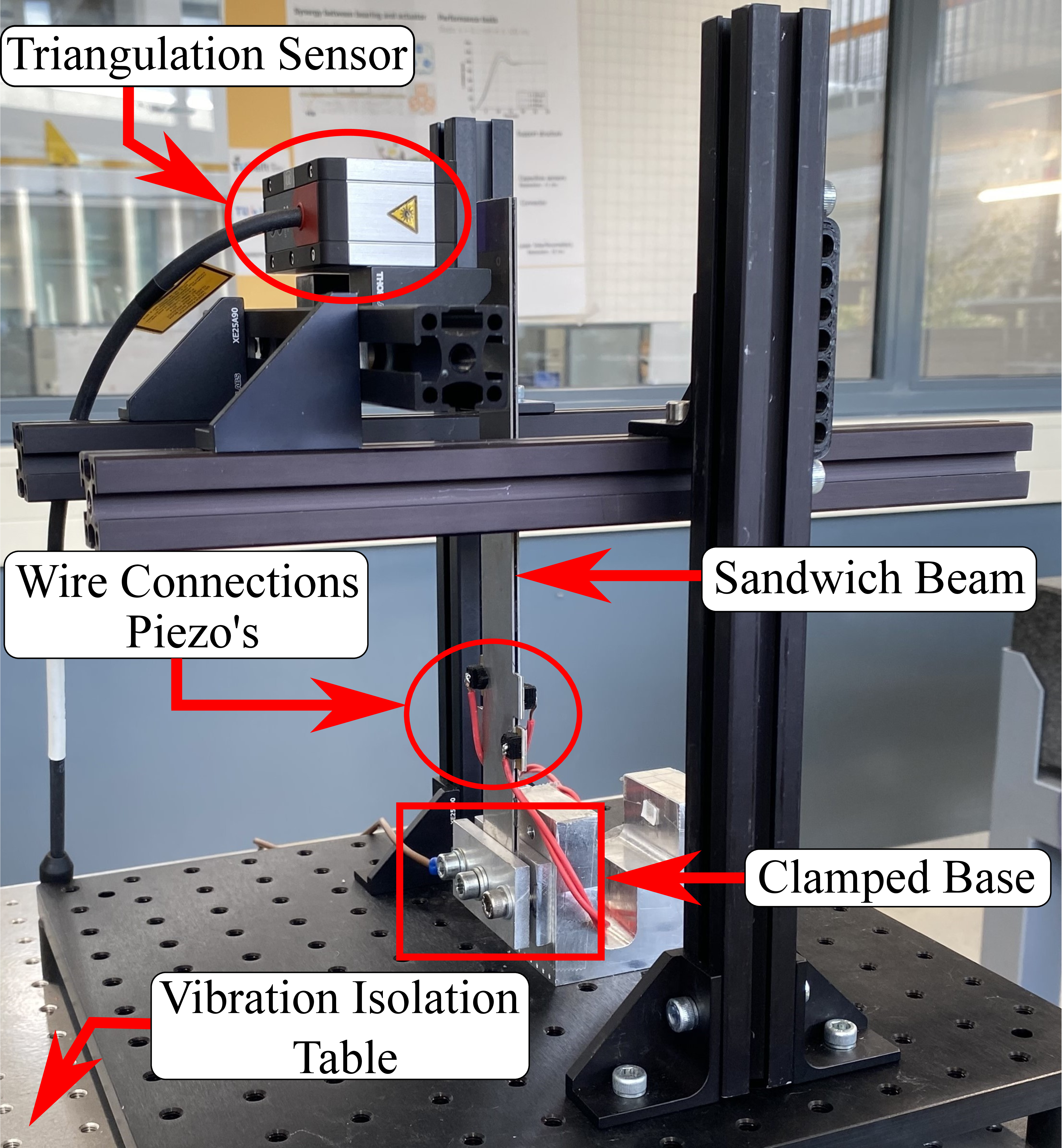} 
		\subcaption{}
		\label{fig:setup_a}
	\end{minipage}
	\hfill
	\begin{minipage}[h]{0.49\textwidth}
		\centering
		\includegraphics[height=7cm]{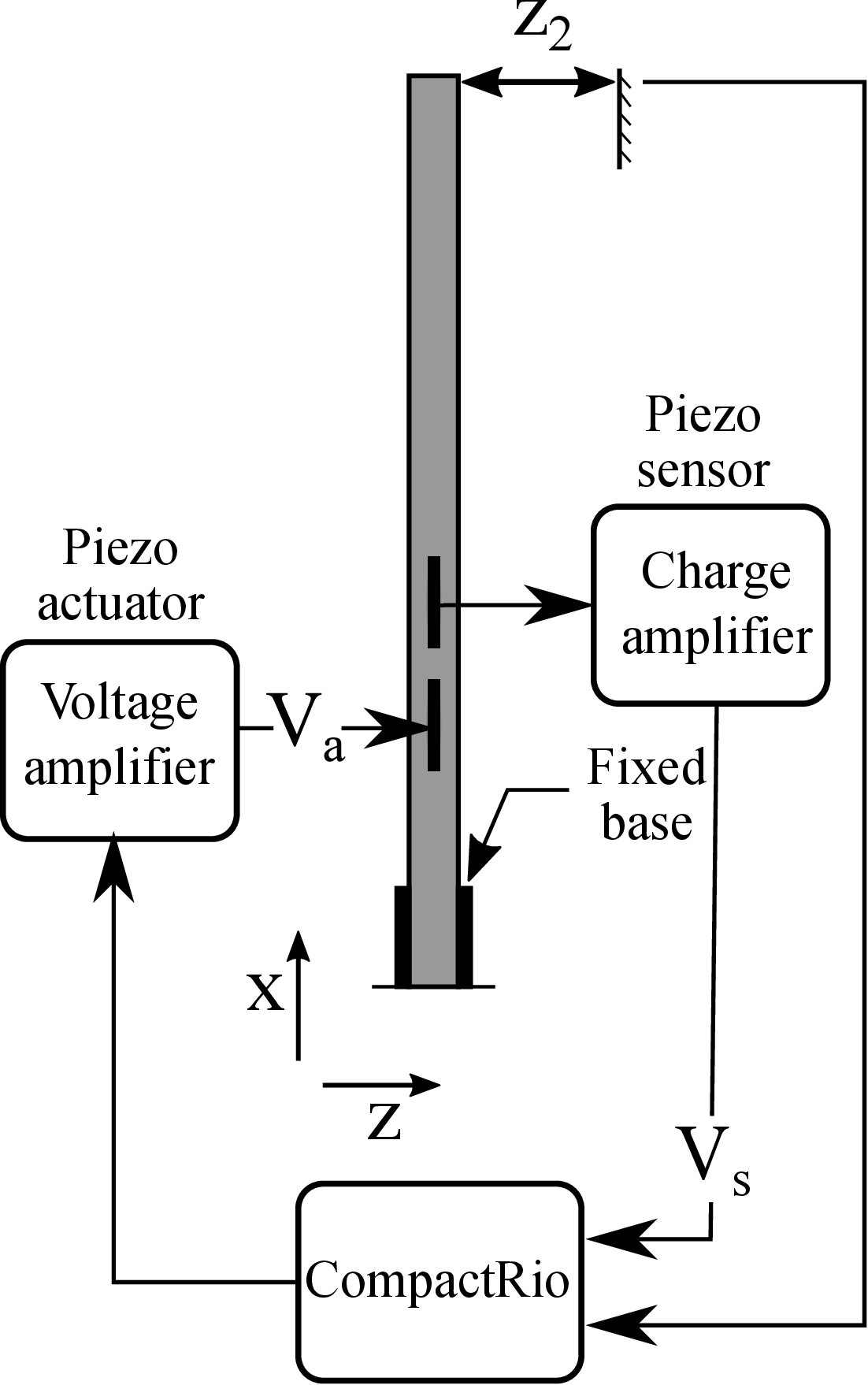}
		\subcaption{}
		\label{fig:setup_b}
	\end{minipage}
	\caption{Test setup used for evaluation and validation of the open-loop response of the sandwich beam structure, utilizing both a piezoelectric shear actuator and sensor. (a) Test setup used to perform experimental system identification. The base of the sandwich structure is clamped around the \textit{Clamped Base} section. The setup is mounted on a \textit{Vibration Isolation Table}, and the tip displacement of the sandwich structure is measured by a \textit{Triangulation Sensor}. (b) Representation of the test setup. $Z_2$ represents the measured tip displacement. A voltage signal is sent by the \textit{CompactRio} controller module to the \textit{Piezo actuator}, and the output signal is amplified by a \textit{Voltage amplifier} $V_a$. The input signal of the \textit{CompactRio} is conditioned by a \textit{Charge amplifier} $V_s$.}
	\label{fig:setup} % Label for the entire figure
\end{figure}

For the open-loop identification, the sandwich beam was mounted by bolts and clamping blocks to the base holder as shown in Fig.\ref{fig:setup_a}
In this way, the base of the sandwich structure was fixed, whereas the end is free, denoting a clamped-free sandwich structure. The whole setup has been isolated from outside disturbances by the \textit{Table-Stable TS-150} vibration isolation table. The piezoelectric actuator requires an electric field intensity of 100-640 $V/mm$ to obtain significant actuation. A high-voltage amplifier \textit{Smart Material HVA 1500/500} with a magnitude gain of 200 $V/V$ has been used. The output voltage of the \textit{CompactRio NIc RIO-9039} controller module was limited by maximum  $\pm$ 1.0 Volt, in order to stay way below the maximum limit of 640 V/mm. The CompactRio was employed with a \textit{NI 9201} 12-bits input module and a \textit{NI 9264} 16-bits output module.  A self-made charge amplifier, based on TL074 operational amplifiers \cite{karki2000signal}, with $C_f = 22nF$ and $R_f = 2M\Omega$ is used to condition the signals of the piezo sensor. The transfer function\cite{kaczmarek2024creating} \cite{hani2021adaptive} between the charge of the sensor and the amplifier output voltage is shown in Eq. \ref{eq:charge_amp}.
\begin{equation} \label{eq:charge_amp}
	\frac{V_s}{Q} = \frac{-R_f s}{(R_f C_f s + 1)(R_i(C_p + C_c)s + 1)}
\end{equation}
Where the low- and high- cut-off frequency are defined by the poles $\omega_1 = 1/(R_f C_f)$ and $\omega_2 = 1/R_i(C_p + C_c)$ respectively. $C_c$ and $R_i$ represent the cable capacitance and input resistance, which were both relatively low, observed by the fact that $f_2$ was found to be above the bandwidth of the experiments. The gain of the flat frequency band equals $1/C_f$. The \textit{Micro-Epsilon Opto NCDT-1900} triangulation sensor was located at the tip of the beam to be able to measure the displacement of the fundamental mode of the beam.

For the open-loop identification, the piezoelectric shear actuator has been used as a disturbance source to excite the modes of the beam structure. A sinusoidal chirp signal was used with a frequency gradually building up from 1 to 1000 Hz. The CompactRio employed a sampling frequency of 10kHz. The system has been identified by evaluating the response of the piezoelectric sensor and the triangulation sensor.
\noindent

\subsection{FEM model: Active Vibration Control}
\begin{figure} [h!]
	\centering
	\begin{tabular}{c}
		\includegraphics[height=5.5cm]{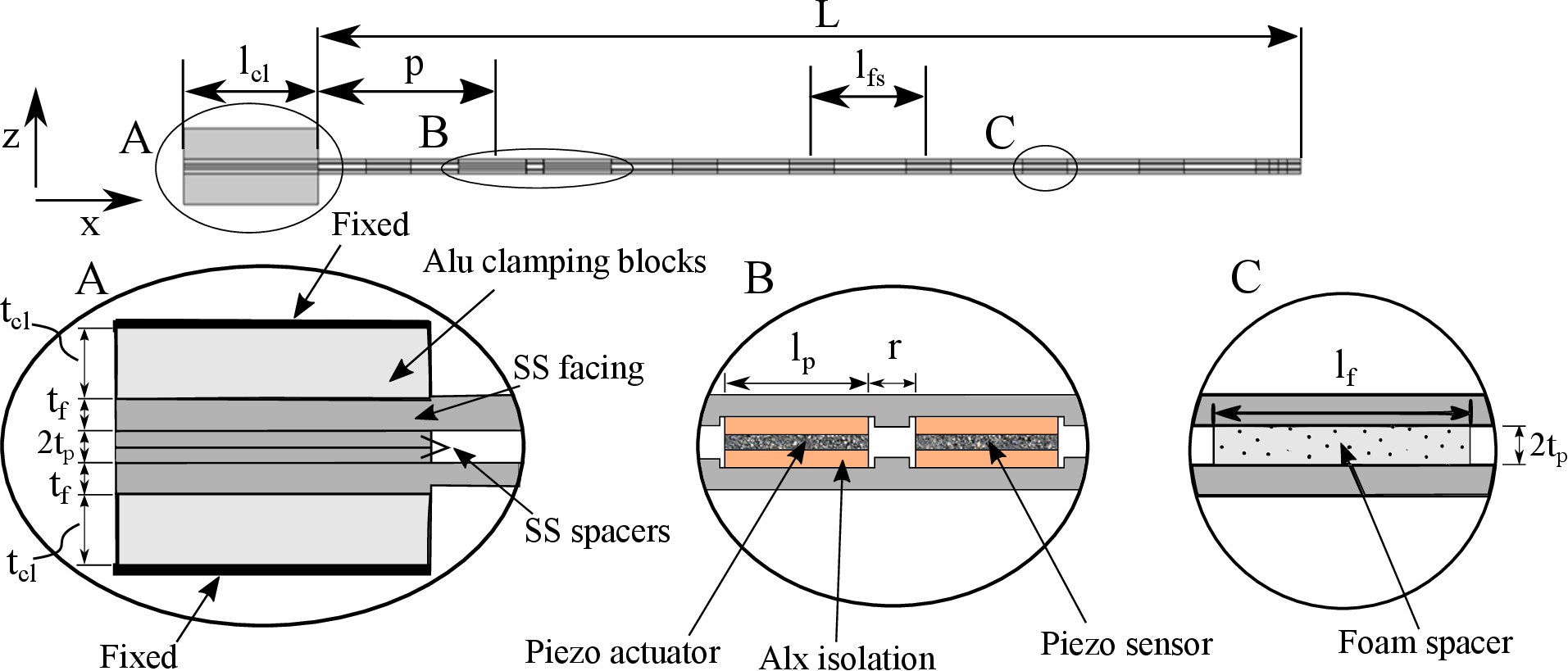}
	\end{tabular}
	\caption[example] 
	%>>>> use \label inside caption to get Fig. number with \ref{}
	{ \label{fig:FEM_model2} 
		Graphical representation of the sandwich beam structure with collocated sensor-actuator pair. Zoomed section A shows the clamped base of the structure, where the aluminum (Al) clamping blocks were fixed at the black bars. Thin glue layers were added in the model between the stainless steel (SS) facings and SS spacers, as well as in, zoom-section B between the piezoelectric transducers, alumina (alx) isolation and SS facings. Dimensions of the piezo actuator and sensor are equal. Parameter values and material constants can be found in \ref{Appendix:B_parameter}
		.}
\end{figure}

This section describes the enhanced 2D FEM model, which has been used to simulate the active damping performance of the sandwich beam structure utilizing both a piezoelectric shear sensor and piezoelectric shear actuator, see Fig. \ref{fig:FEM_model2}. This model has been verified according to experimental open-loop evaluation. 

\subsubsection{Open-loop system-identification}
First the open-loop frequency response has been evaluated by the numerical model. Similar boundary conditions as for the experimental test (see Fig. \ref{fig:setup_b}) were employed in this simulation, except for the fixed base. In the simulation the fixed base has one degree of freedom and was excited with a disturbance acceleration in $z-$direction, denoted with $\ddot{z_1}$. The identification has been performed by evaluating the harmonic response of the structure from $\ddot{z_1}$ and $V_a$ to both $z_2$ and $V_s$. Accordingly, the following Multiple-Input Multiple-Output (MIMO) system has been obtained, where $\ddot{z_1}$ and $V_a$ define the inputs and $z_2$ and $V_s$ the outputs of the MIMO system. 
\begin{equation} \label{eq:MIMO}
	\left[
	\begin{array}{c}
		\mathrm{z}_2 \\
		\mathrm{V}_s	 
	\end{array}
	\right]
	=
	\left[
	\begin{array}{cc}
		H_{11} & H_{12} \\
		H_{21} & H_{22} 
	\end{array}
	\right]
	\left[
	\begin{array}{c}
		\ddot{\mathrm{z}}_1 \\
		\mathrm{V}_a
	\end{array}
	\right]
\end{equation}
Where $H_{11}$, $H_{12}$, $H_{21}$ and $H_{22}$ represent the frequency response of the system for $(z_2/\ddot{z_1})$,  $(z_2/V_a)$, $(V_s/\ddot{z_1})$ and $(V_s/V_a)$ respectively. The input voltage $V_a$ of the piezoelectric actuator can be derived from the constitutive equations of Eq. \ref{eq:const_eq} by Eq. \ref{eq:Va}.
\begin{equation} \label{eq:Va}
	V_a = E_1 t_p
\end{equation}
\noindent
Where $t_p$ is the thickness of the piezoelectric transducer. In the experimental section, the piezo sensor had been connected to a charge amplifier, therefore the sensor was simulated in short-circuit conditions. Accordingly, the output voltage of the piezoelectric sensor $V_s$ is related to the electric charge displacement by Eq. \ref{eq:Vs}.
\begin{equation} \label{eq:Vs}
	V_s = \frac{D_1 l_p b_p}{C_f}
\end{equation}
\noindent
Where $D_1$, $l_p$ and $b_p$ are the electric charge displacement, length and width of the transducer respectively. Recall that $1/C_f$ represents the gain of the flat frequency band of the charge amplifier employed in the experimental work. 
\subsubsection{Architecture active vibration controller}
\begin{figure} [h!]
	\centering
	\begin{tabular}{c}
		\includegraphics[height=2.5cm]{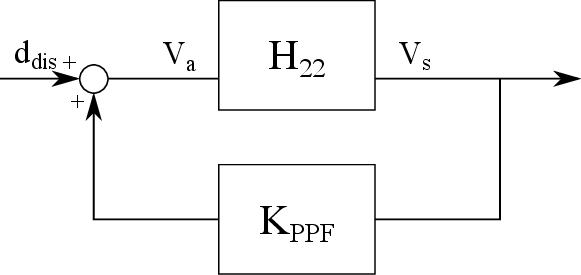}
	\end{tabular}
	\caption[example] 
	%>>>> use \label inside caption to get Fig. number with \ref{}
	{ \label{fig:control_scheme} 
		Control scheme of the closed loop system. $H_{22}$ refers to the frequency response of $\frac{V_s}{V_a}$. $K_{ppf}$ represents the PPF-controller. Note that the closed loop utilizes positive feedback.}
\end{figure}

The active vibration controller utilized in this paper is Positive Position Feedback (PPF). This controller was first introduced by Goh \& Caughey \cite{fanson1990positive}. The PPF controller uses position measurement to gain damping, which makes it perfectly suitable for the position based piezo sensor used in this paper. Advantages of PPF control are that it avoids differentiation of the signal and creates a high gain roll-off after the cut-off frequency of the controller, both these effects reduce destabilization by (unknown) uncontrolled higher order (residual) modes of the structure \cite{mao2013control}. The controller is similar to a second order low-pass filter except for an adjusted gain, see Eq. \ref{eq:PPF}.  
\begin{equation} \label{eq:PPF}
	K_{ppf} = \frac{H_{0}^{-1} \, g \, \omega_f^2}{s^2+2\zeta_f \omega_f s + \omega_f^2}
\end{equation} 
\noindent
$H_0$ represents the low frequency gain of $H_{22}$, $g$ the PPF control gain, $\zeta_f$ the PPF damping term and $\omega_f$ the controller cut-off frequency. To generate damping by the actuator and sensor employed in this system, the controller is designed on the transfer ($V_s/V_a)$, targeting the fundamental pole. The block diagram of the feedback system is shown in figure \ref{fig:control_scheme}.
\noindent
To dampen the tip vibrations of the sandwich beam structure, the closed-loop damping introduced by the PPF-controller on the tip-displacement has been evaluated by Eq. \ref{eq:z1_z2}. This system can be obtained by implementing the PPF-controller from Eq. \ref{eq:PPF} to the transfer ($V_s/V_a)$ (as shown in Fig. \ref{fig:control_scheme}) in the MIMO system of Eq. \ref{eq:MIMO}.

\begin{equation} \label{eq:z1_z2}
	\frac{z_2}{\ddot{z_1}} = H_{11} - H_{12} \cdot \frac{K_{ppf}}{1 + H_{22} \cdot K_{ppf}} \cdot H_{21}
\end{equation}
\noindent

\section{Results \& discussions}
 
\subsection{Location piezoelectric transducer}
\label{sec:optimal_x}
\begin{figure}[h!]
	\begin{subfigure}[h]{0.49\textwidth}
		\centering
		\begin{tabular}{c}
			\includegraphics[height=5.5cm]{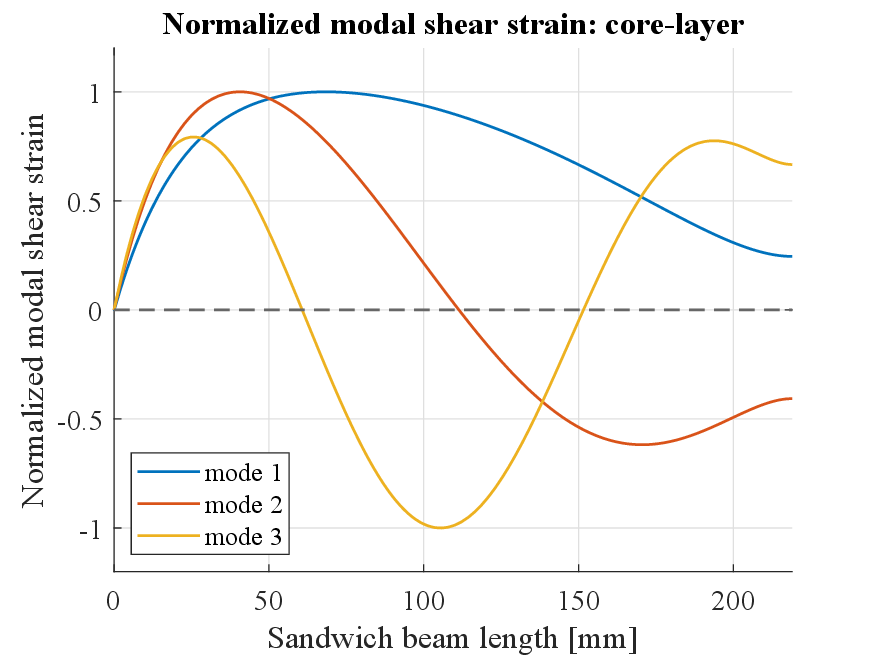} 
		\end{tabular}
		\subcaption{} 
		\label{fig:res_optx_a}
	\end{subfigure}
	\hfil
	\begin{subfigure}[h]{0.49\textwidth}
		\centering
		\begin{tabular}{c}
			\includegraphics[height=5.5cm]{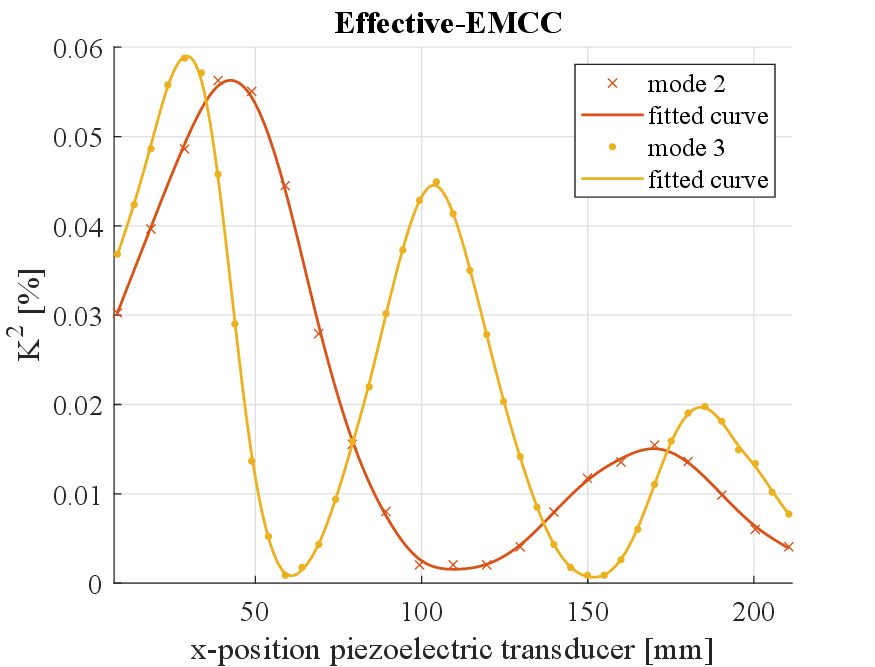}
		\end{tabular}
		\subcaption{}
		\label{fig:res_optx_b}
	\end{subfigure}
	\caption[example] 
	%>>>> use \label inside caption to get Fig. number with \ref{}
	{ \label{fig:res_optx} Results according to the normalized modal shear strain and Effective-EMCC. (a) Representation of the normalized modal shear strain of the core-layer for the first three bending modes. The horizontal axis defines the $x-$location in the sandwich beam in mm, where $x = 0$ refers to the clamped base and $x = 219$ to the free end of the structure. (b) Representation of the Effective-EMCC for bending mode 2 and 3. The smooth spline fitted curve shows the expected trend based on the plotted data points. The horizontal axis defines the x-location of the piezoelectric transducer inside the sandwich beam structure in mm, where $x = 0$ refers to the clamped base and $x = 219$ to the free end of the structure}
\end{figure}

The $x-$location refers to the location of the piezoelectric shear transducer in the length direction of the sandwich beam structure. As mentioned in the previous section, the location has been evaluated according to two methods: (1) normalized modal shear strain and (2) the Effective-EMCC. In Fig. \ref{fig:res_optx_a} the normalized modal shear strain for bending mode 1,2 and 3 are shown, where the strain has been normalized with respect to the mode under study. The shear strain normalized with respect to mode 3 is shown in Fig. \ref{fig:rel_strain} in appendix \ref{Appendix:strain}. This clearly shows the relative difference between the strain of mode 1, 2 and 3. In Fig. \ref{fig:res_optx_b} the EMCC for bending mode 2 and 3 of the sandwich beam structure are represented.
\\
The first mode is excluded within the EMCC, since the difference between the open- and short-circuit conditions was negligible and therefore the resolution limit of the simulation was reached. Therefore, $K_1^2$ tends to be relatively low compared to the EMCC of higher order modes 2 and 3. Taking this into account when evaluating Fig. \ref{fig:res_optx_b}, one can observe that the EMCC increases for the higher order mode, but overall the EMCC is relatively low. Similar observations were noticed for other sandwich beam structures \cite{article}. In the study of Trindade et. al. it was suggested that the increase of the EMCC was being caused by the relative increase of the modal shear strain for higher order modes. In this study the increase of $K_3^2$ compared to $K_2^2$ can be confirmed by Fig. \ref{fig:rel_strain}, which clearly shows the relative increase for the modal shear strain of mode 3 and 2 compared to mode 1. 
\noindent
The EMCC and normalized modal shear strain show correspondence in the $x-$location of the maxima and zero values for mode 2 and 3, except for the decay of $K_i^2$ towards the free-end of the sandwich beam structure. The observed difference between the normalized shear strain and EMCC might be related to the modal dynamics of the shear piezo, which are only present in the EMCC analysis. This can be explained by the fact that the shear piezo has relatively high stiffness ($1/s_{55} = 22.2 \; GPa$) and density ($\rho_p = 7750 \; kg/m^3$) compared to the soft ($G_{fc} = 13.5 \; MPa$) and light-weighted ($\rho_{fc} = 35 \; kg/m^3$) core layer. In this way, the shear piezo will increase the mass and stiffness of the structure locally, which will effect the overall modal dynamics of the sandwich structure. Therefore, the EMCC analysis is found to be superior to the modal shear strain analysis. 
\\
\\
The maxima for the EMCC of mode 2 has been used in this study to define the $x-$location (as shown in Fig. \ref{fig:FEM_model2}, $p = 38.8 \; mm$) of the piezoelectric actuator. This has three main reasons: (1) the EMCC includes the modal dynamics of the piezo and therefore these results tend to be more realistic compared to the normalized modal shear strain. (2) In this study the actuator and sensor are placed in series inside the core-layer of the sandwich structure. Since, the actuator should counter-act the strains, the actuator should be located exactly at the maxima to achieve maximum performance. The sensor should only be capable of measuring the strains induced by the actuator and the modal vibrations of the structure. A control gain can be used to compensate for the difference in amplitude between what the sensor is measuring and the actuator is demanding. Therefore, the sensor should be placed closely to the actuator to achieve good coupling between actuator and sensor ((nearly) collocated), but should not necessarily be placed exactly at the maximum modal strain/EMCC. (3) The optimal location according to the maximum Effective-EMCC for the second mode partly covers the optimal region for the Effective-EMCC of mode three and the modal shear strain of mode one.

\subsection{Open-loop identification}
\begin{figure}[h!]
	\begin{subfigure}[h]{0.49\textwidth}
		\centering
		\begin{tabular}{c}
			\includegraphics[height=5.4cm]{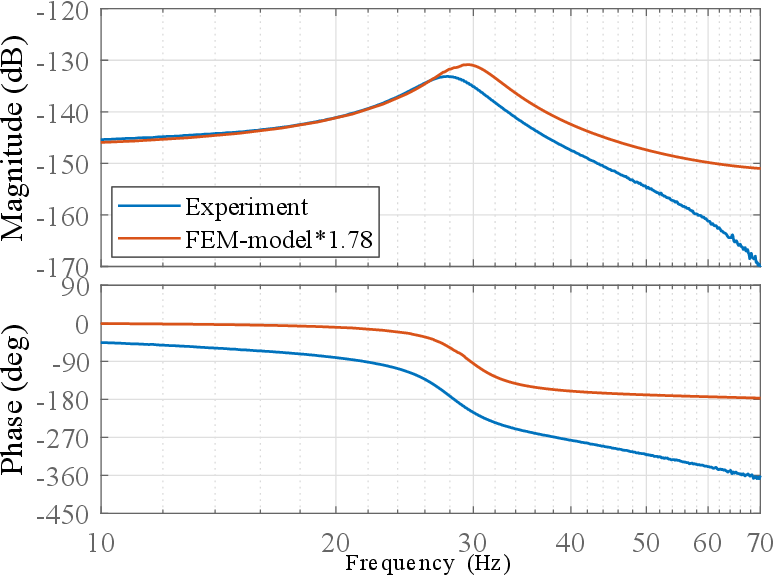}
		\end{tabular}
		\subcaption{}
		\label{fig:exp_model_a}
	\end{subfigure}
	\hfill
	\begin{subfigure}[h]{0.49\textwidth}
		\centering
		\begin{tabular}{c}
			\includegraphics[height=5.4cm]{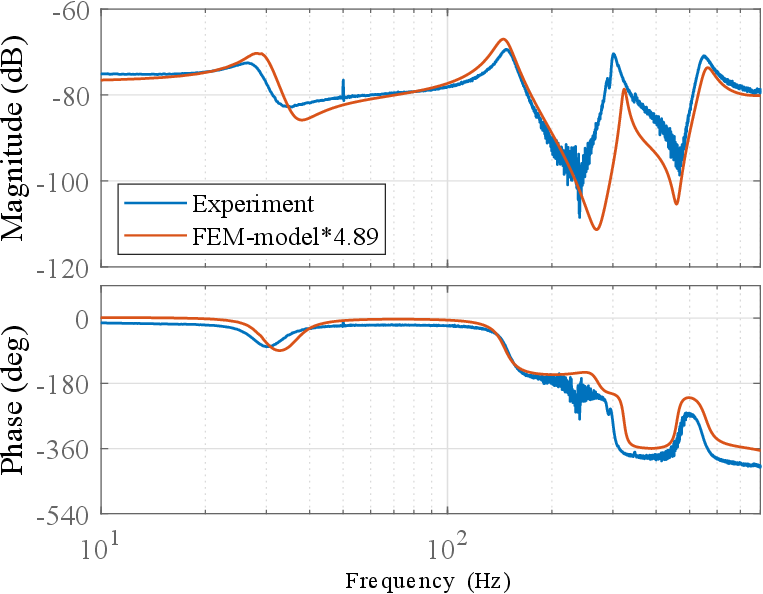}
		\end{tabular}
		\subcaption{}
		\label{fig:exp_model_b}
	\end{subfigure}
	\caption[example] 
	%>>>> use \label inside caption to get Fig. number with \ref{}
	{ \label{fig:exp_model} Comparison between the frequency response of the experimental work and the 2D FEM model. (a) Open-loop comparison of $\frac{z_2}{V_a} \; [m/V]$. Gain of FEM-model is multiplied by 1.78 to match the gain of both responses. (b) Open-loop comparison of $\frac{V_s}{V_a} \; [$V/V$]$. Gain of FEM-model is multiplied by 4.89 to match the gain of both responses. By the pole-zero flipping the phase crosses -180\textdegree.}
\end{figure}

In this subsection the response of the FEM model is compared to the response obtained in the experimental work. In Fig. \ref{fig:exp_model} the frequency response of both systems are shown. Fig. \ref{fig:exp_model_a} shows the response for actuator to tip-displacement ($z_2/V_a$)  and Fig. \ref{fig:exp_model_b} the response for actuator to sensor ($V_s/V_a$). During the practical work delamination was observed between both piezoelectric transducers and the structure. Therefore, the glue layer stiffness has been reduced for the connecting layers of the piezo's, alumina layers and facings in the FEM model. This has been done by reducing the stiffness by 10 \% and 0.5 \% of its original stiffness at the layers concerning the actuator and sensor region respectively. 
\\
\\
From Fig. \ref{fig:exp_model_a} it can be clearly observed that the sandwich beam structure has a relatively high inherent structural damping. This is probably caused by the damping introduced by the foam spacers, since foam materials have a relative high mechanical loss coefficient [0.005, 0.5]\cite{grantaedupack} compared to stainless steel [3.1e-4, 1.2e-3]\cite{grantaedupack} and alumina [1e-5, 2e-5]. \cite{grantaedupack} The damping of the experimental work had been evaluated by the half-power bandwidth method \cite{de2006vibration}, where the structural damping of the model was matched accordingly. For the fundamental mode a damping factor of $\zeta^{struct} =  0.11$ has been estimated. The response of the FEM model and experimental work match sufficient. Interesting to note is that the phase is crossing -180\textdegree \- after the second resonance mode, see Fig. \ref{fig:exp_model_b}. This effect is caused by pole-zero flipping, where the second zero (right after the second mode) is followed by another zero instead of a pole. Therefore, this system defines a nearly collocated system \cite{preumont_vibration_2011}, since it contains the interlacing pole-zero property for the fundamental mode only. However, one may observe a different response for systems with perfectly bonded shear piezo elements. 

\subsection{Active Vibration Control}
\begin{figure}[t!]
	\centering
	\begin{minipage}[h]{0.49\textwidth}
		\centering
		\includegraphics[height=5.4cm]{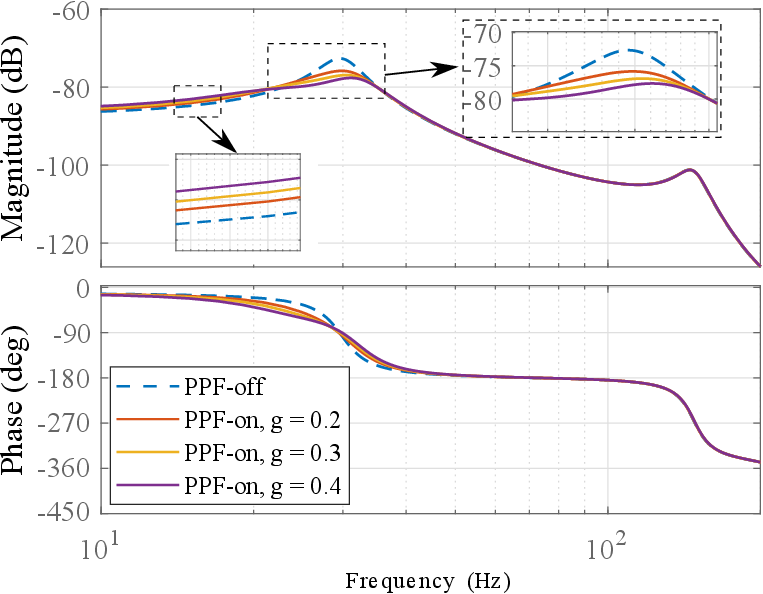}
		\subcaption{} 
		\label{fig:PPF_damped_a}
	\end{minipage}
	\hfill
	\begin{minipage}[h]{0.49\textwidth}
		\centering
		\includegraphics[height=5.4cm]{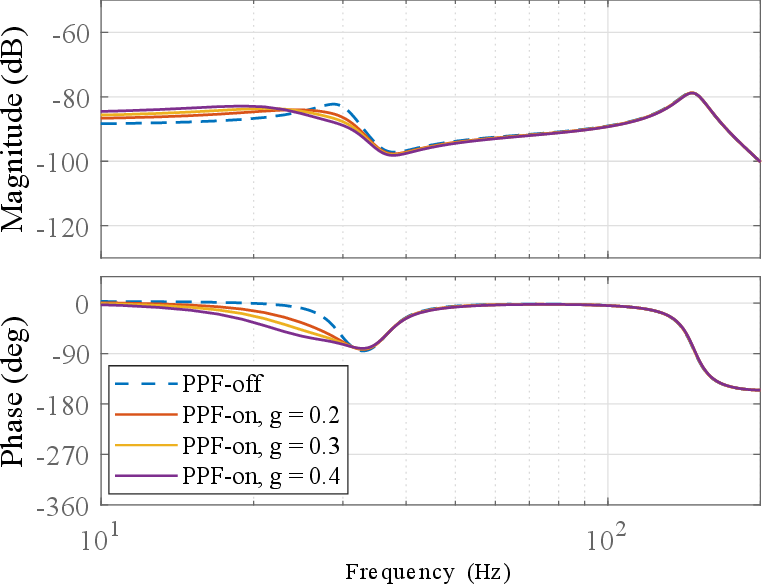}
		\subcaption{}
		\label{fig:PPF_damped_b}
	\end{minipage}
	\caption{Frequency response of the closed-loop system for different control gains. (a) Closed-loop comparison of $\frac{z_2}{\ddot z_1} \; [m/(m/s^2)]$ for when PPF is turned off and on with different control gains. Higher control gains result in better resonance suppression, despite resulting in unwanted increases in the low-frequency gain. (b) Closed-loop comparison of $\frac{V_s}{V_a} \; [$V/V$]$ for when PPF is turned off and on with different control gains. Due to high inherent structural damping, the effect of the PPF controller is limited.}
	\label{fig:PPF_damped} % Label for the entire figure
\end{figure}

PPF control has been used to examine the active damping performance of the sandwich beam utilizing both a piezoelectric shear sensor and piezoelectric shear actuator. First, the performance of the controller is examined by evaluating the frequency response for different control gains. Fig. \ref{fig:PPF_damped_b} shows the effect of the PPF controller on the closed-loop response from actuator to sensor ($V_s/ V_a$). Fig. \ref{fig:PPF_damped_a} shows the closed-loop response of the system from base-disturbance to tip-displacement ($z_2/ \ddot{z_1}$). The controller cut-off frequency has been set equal to the fundamental mode.
The damping term $\zeta_f$ has been defined iteratively and finally set equal to $\zeta_f = 0.3$. This $\zeta_f$ value is commonly used in practice, since it incorporates the fact that it is difficult to accurately tune the PPF controller to exactly/only target the resonance region \cite{kwak2022dynamic}. The nyquist diagram of the open-loop system is shown in appendix \ref{Appendix:C_nyquist}.

A well known characteristic of PPF-control is that it increases the low frequency gain of for higher control gains. This effect is also clearly visible in Fig. \ref{fig:PPF_damped_a}. Depending on the area of interest, the choice for the control gain value is always a trade-off between good suppression of the resonance region and increase of the low frequency gain. For a control gain of $g = 0.3$, the low frequency gain has been increased with only 1.17 dB at 15 Hz. 
Due to the high structural damping, the PPF controller adds a limited amount of extra damping to the closed loop system. The amount of damping from base-disturbance to tip-displacement is limited by a relative reduction of 4.4 dB at fundamental mode for g = 0.3 (see Fig. \ref{fig:PPF_damped_a}). 
Therefore, the structural damping of the system has been lowered to $\zeta^{struct} = 0.01$ in the FEM model. This was done to make the system more comparable to other active damping applications considering beam structures with a lower structural damping\cite{ajjaj2022vibration}. Again, the controller cut-off frequency has been set equal to the fundamental mode and the PPF damping term $\zeta_f$ has been found iteratively and set equal to $\zeta_f = 0.3$. The nyquist diagram of the open-loop system is shown in appendix \ref{Appendix:C_nyquist}. In Fig. \ref{fig:PPF_lowdamp}, the frequency response of the system for different controller gains is represented.   
\begin{figure}[t!]
	\centering
	\begin{minipage}[h]{0.49\textwidth}
		\centering
		\includegraphics[height=5.4cm]{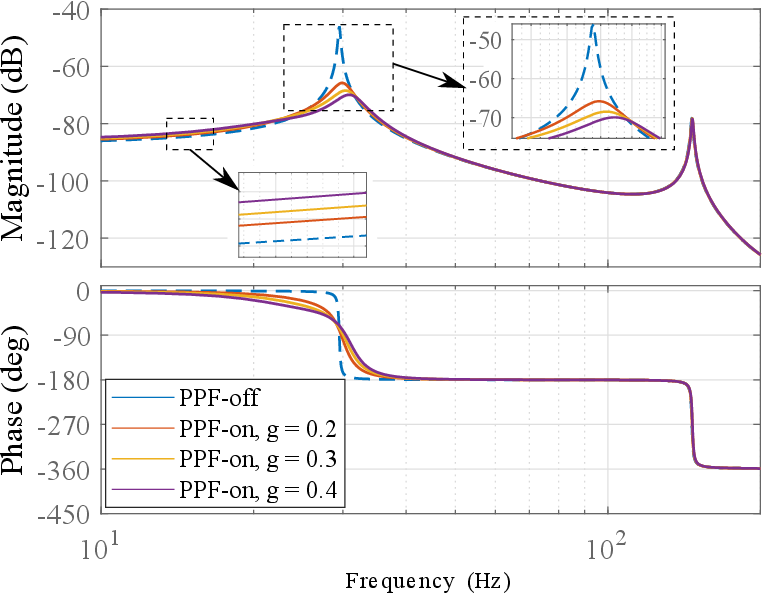}
		\subcaption{} 
		\label{fig:PPF_lowdamp_a}
	\end{minipage}
	\hfill
	\begin{minipage}[h]{0.49\textwidth}
		\centering
		\includegraphics[height=5.4cm]{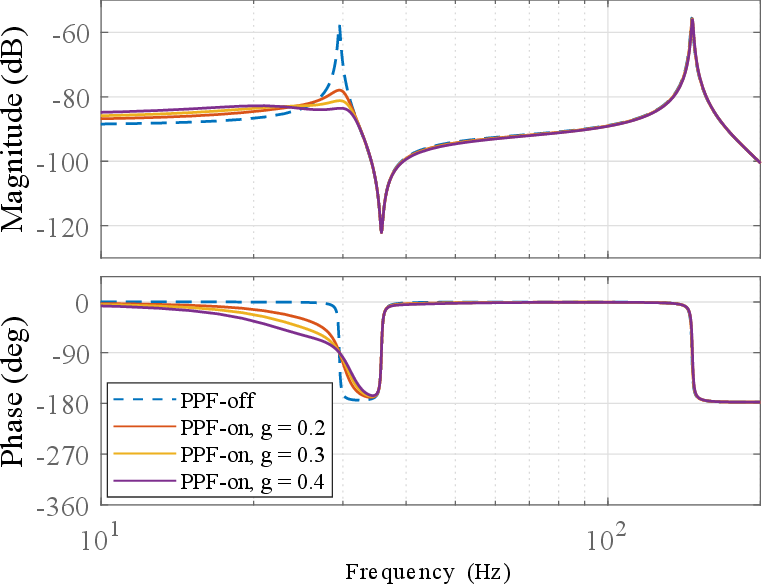}
		\subcaption{}
		\label{fig:PPF_lowdamp_b}
	\end{minipage}
	\caption{Frequency response of the closed-loop system with lowered inherent structural damping. The response of the system is evaluated for different control gains. (a) Closed-loop comparison of $\frac{z_2}{\ddot z_1} \; [m/(m/s^2)]$ when PPF is turned off and on with different control gains. Higher control gains result in better resonance suppression, despite resulting in unwanted increases in the low-frequency gain. (b) Closed-loop comparison of $\frac{V_s}{V_a} \; [$V/V$]$ for when PPF is turned off and on with different control gains. Due to the lowered inherent structural damping, the PPF controller introduces significant damping.}
	\label{fig:PPF_lowdamp} % Label for the entire figure
\end{figure}
From Fig. \ref{fig:PPF_lowdamp_b} one can clearly observe that the PPF controller now introduces significant damping compared to when the controller is being turned off. Fig. \ref{fig:PPF_lowdamp_a} shows a relative reduction of 22.8 dB (for $g = 0.3$) for the fundamental mode. For $g= 0.3$, the low frequency gain has been increased by only 1.13 dB at 15 Hz.

\begin{figure}[ht!]
	\begin{subfigure}[h]{0.49\textwidth}
		\centering
		\begin{tabular}{c}
			\includegraphics[height=5.4cm]{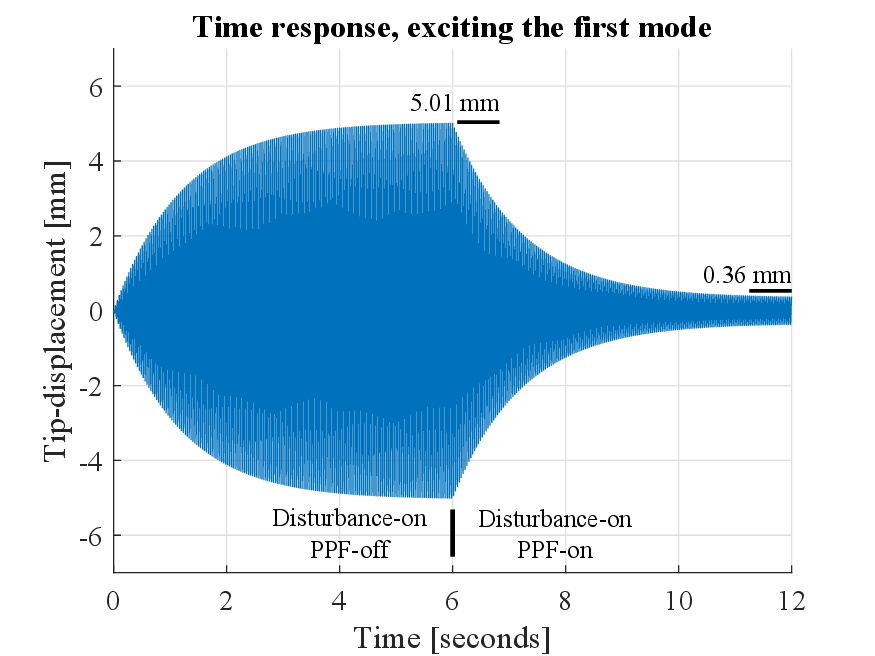}
		\end{tabular}
		\subcaption{} 
		\label{fig:time_resp_a}
	\end{subfigure}
	\hfill
	\begin{subfigure}[h]{0.49\textwidth}
		\centering
		\begin{tabular}{c}
			\includegraphics[height=5.4cm]{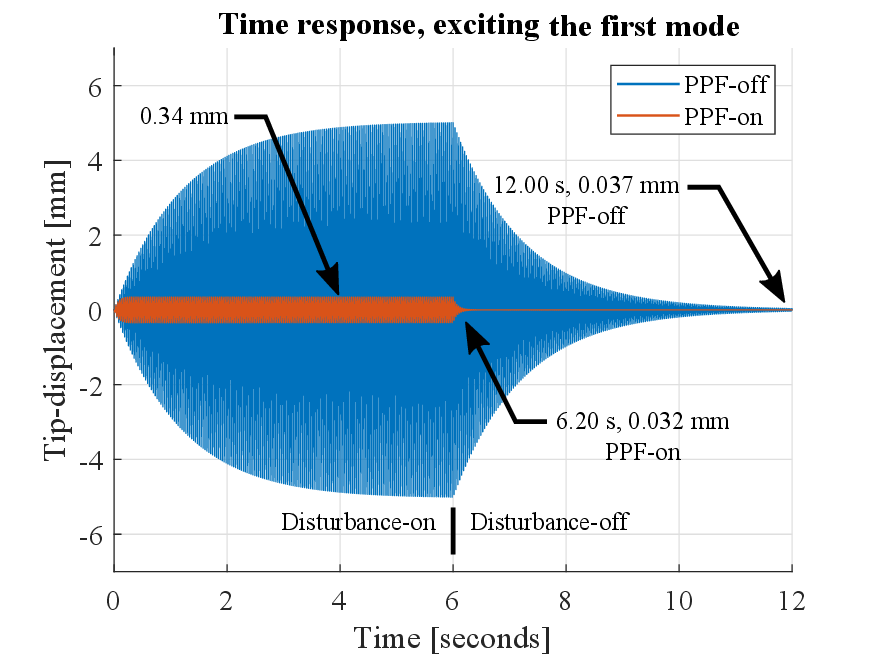}
		\end{tabular}
		\caption{}
		\label{fig:time_resp_b}
	\end{subfigure}
	\caption[example] 
	%>>>> use \label inside caption to get Fig. number with \ref{}
	{ \label{fig:time_resp} Time response simulations of the tip-displacement by exciting the resonance frequency of the first mode with a sinusoidal base-disturbance signal of 1 $[m/s^2]$ amplitude. (a) Tip-displacement when the PPF control is switched on after 6 seconds. Amplitude reduces from 5.01 mm at time = 6.0 seconds to 0.36 mm at time = 12.0 seconds. (b) Settling time of the tip vibrations. Disturbance signal is switched off after time = 6.0 seconds. For no PPF control tip displacement reduces to 0.037 mm amplitude at time = 12.0 seconds. When PPF control is switched on the tip vibrations settle within a bound of 0.032 mm amplitude within 0.20 seconds. When PPF is on and disturbance is on, the tip-vibrations stabilize to 0.34 mm amplitude. } 
\end{figure}

The system of Fig. \ref{fig:PPF_lowdamp} has been used to simulate the time response of the system from base-disturbance to tip-displacement ($z_2/ \ddot{z_1}$). For this simulation the PPF control gain was set equal to $g = 0.3$. A sinusoidal signal was employed as base-disturbance signal. Where the frequency has been set equal to the resonance frequency of the fundamental mode $f_1 = 29.5$ Hz with an amplitude of 1 $m/s^2$. The time response of the tip-displacement is shown in Fig. \ref{fig:time_resp_a}, where the PPF controller has been turned on after 6 seconds. Fig. \ref{fig:time_resp_b} shows the difference between the time-response of the tip-displacement when the PPF controller is switched on and off. In addition, the disturbance signal has been turned off after time equals 6.0 seconds, to indicate the settling time for PPF-on and PPF-off. 
\noindent
From Fig. \ref{fig:time_resp_a} the impact of the controller on the tip-displacement is clearly visible, where the amplitude is reduced from 5.01 mm at time equals 6.0 seconds to 0.36 mm at time equals 12.0 seconds. From Fig. \ref{fig:time_resp_b} one can observe that the tip vibrations stabilize to 0.34 mm amplitude when PPF is on and disturbance is on. When disturbance is off and the PPF controller is off, the amplitude of the tip vibrations reduce to 0.037 mm within 6.0 second. However, when the PPF is on and disturbance is on the tip-vibrations settle within a bound of $\pm$ 0.032 mm within 0.20 seconds, indicating a significant reduction with a settling time which is at least 30 times faster.

\section{Conclusion}
\begin{itemize}[noitemsep]
	\item In this study the damping performance for long and slender sandwich beam structures by an internally placed piezoelectric shear actuator and piezoelectric shear sensor has been demonstrated by utilizing Positive Position Feedback control. 
	\item The optimal location of the piezoelectric transducer is found at the maximum Effective-EMCC for the second mode, which partly covered the optimal region for the Effective-EMCC of mode three and the modal shear strain of mode one. 
	\item The Effective-EMCC analysis is found to be superior to the modal shear strain analysis since it incorporates the dynamics of the piezo element. 
	\item Both the modal shear strain of the core-layer and Effective-EMCC increase for higher order modes.  
	\item A sandwich beam structure consisting of both a piezoelectric shear actuator and sensor has been fabricated to perform open-loop experiments.
	\item Good correspondence was found between the open-loop identification of the experimental work and the enhanced FEM model.
	\item PPF control is successfully implemented on the numerical frequency response. Different control gains have been employed to simulate the trade-off between good resonance suppression and increase of the low frequency gain. 
	\item The closed-loop simulations show that the fundamental mode can be suppressed successfully by 22.8 dB, while increasing the low frequency gain with only 1.13 dB. This caused the tip vibrations at the fundamental mode to decay from 5.01 mm amplitude to 0.34 mm at steady state. 
	\item The proposed solution laid a foundation for (high-tech) systems for suppressing mechanical vibrations in order to enhance the positioning, accuracy and speed, while: (1) addressing the constraints on the design volume and (2) protecting vulnerable piezoelectric sensors and actuators from environmental conditions.
\end{itemize}

\section{Data availability statement}
The data cannot be made publicly available upon publication because they are not available in a format that is sufficiently accessible or reusable by other researchers. The data that support the findings of this study are available upon reasonable request from the authors.

\section{Acknowledgment}
This document is part of the project Metamech (with project
number 17976) of the research programme Applied and
Engineering Sciences which is (partly) financed by the Dutch
Research Council (NWO).

\section*{References}
\bibliographystyle{iopart-num}
\bibliography{bibliography_thesis}
\appendix
\section{Normalized modal shear strain w.r.t mode 3}
\label{Appendix:strain}

\begin{figure} [H]
	\centering
	\begin{tabular}{c}
		\includegraphics[height=6cm]{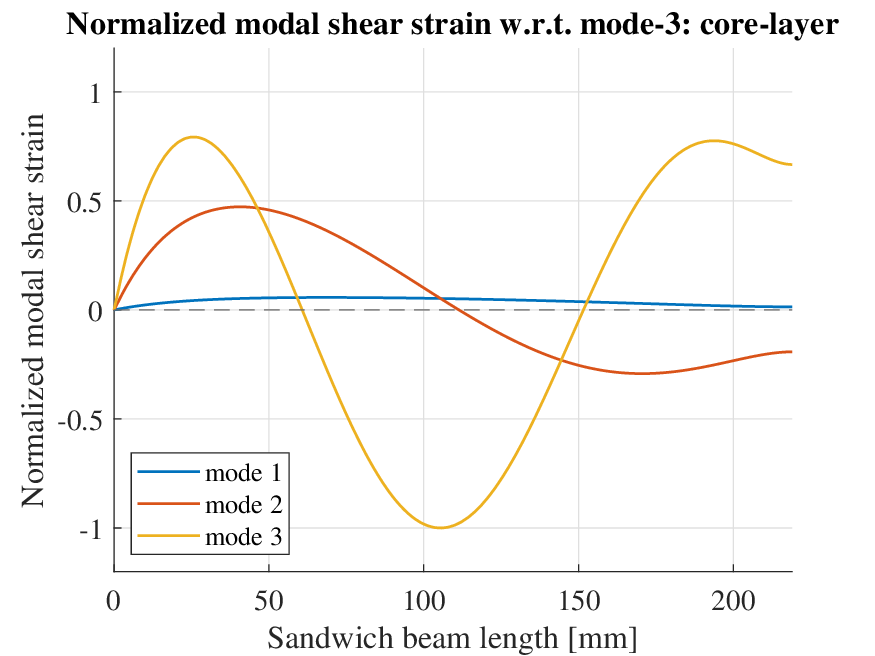}
	\end{tabular}
	\caption[example] 
	%>>>> use \label inside caption to get Fig. number with \ref{}
	{ \label{fig:rel_strain} 
		Normalized modal shear strain of the core-layer with respect to mode 3, indicating the relative difference for the shear strain between the modes. Note that the modal shear strain increases for higher order modes.}
\end{figure}

\section{Piezoelectric material constants}
\label{Appendix:A_piezo}
\begin{table}[H]
	\caption{Piezoelectric material constants.} 
	\label{tab:Paper Margins}
	\begin{center}  
 \scalebox{0.8}{
		\begin{tabular}{l|l|l} 
			\hline
			\textbf{Constant name} & \textbf{Value, unit} & \textbf{Comment}  \\
			\hline
			\hline
			${s_{55}}^E$ & 4.51e-11 $[m^2/N]$ & compliance, under constant electric field\\
			
			 $d_{15}$ & 5.57e-10 $[m/V]$ & piezoelectric constant \\
			
			 $\epsilon^T$ & 1.60e-08 $[F/m]$ & permitivity, under constant mechanical stress\\
			 $\rho_p$ & 7750 $[kg/m^3]$ & density\\
			\hline 
		\end{tabular}
  }
	\end{center}
\end{table}
\section{Sandwich beam parameter \& material constants}
\label{Appendix:B_parameter}
\begin{table}[H]
	\caption{Parameter values.} 
	\label{tab:Paper Margins}
	\begin{center}  
 \scalebox{0.8}{
		\begin{tabular}{l|l|l} 
			\hline
			\textbf{Parameter} & \textbf{Value, unit} & \textbf{Comment}  \\
			\hline
			\hline
			$L$ & 219 $[mm]$ & length sandwich beam\\
			$W$ & 25 $[mm]$ & width sandwich beam\\
			$t_f$ & 1.0 $[mm]$ & thickness facing layer \\
			$t_p$ & 0.5 $[mm]$ & thickness piezo, foam layer and alumina layers \\
			$l_p$ & 15 $[mm]$ &  length piezo\\
			$b_p$ & 25, 35 $[mm]$ &  width piezo simulations, width piezo experiments (including recesses)\\
			$p$ & 38.8 $[mm]$ & distance from base to mid-section piezo actuator\\
			$l_{fs}$ & 26 $[mm]$ & distance between foam spacers, from mid- to mid-section\\
			$l_f$ & 10 $[mm]$ & length foam spacer \\
			$r$ & 4.0 $[mm]$ & length spacing between piezo actuator and sensor\\
			$l_{cl}$ & 30 $[mm]$ & length aluminum clamping blocks \\
			$t_{cl}$ & 6 $[mm]$ & thickness aluminum clamping blocks \\
			$t_{g}$ & 105 $[\mu m]$ & thickness epoxy glue layer, estimated by literature \cite{mustapha2015bonding}\\
			\hline 
		\end{tabular}
  }
	\end{center}
\end{table}
\begin{table}[H]
	\caption{Sandwich beam material constants.} 
	\label{tab:Paper Margins}
	\begin{center}     
 \scalebox{0.8}{
		\begin{tabular}{l|l|l} 
			\hline
			\textbf{Parameter name} & \textbf{Value, unit} & \textbf{Comment}  \\
			\hline
			\hline
			$E_{ss}$ & 197 $[GPa]$ & young's modulus stainless steel\cite{grantaedupack}\\
			$\nu_{ss}$ & 0.29 $[-]$ & poisson's ratio  stainless steel\cite{grantaedupack}\\
			$\rho_{ss}$ & 8000 $[kg/m^3]$ & density stainless steel\cite{grantaedupack}\\
			$E_{alu}$ & 2700 $[GPa]$ & young's modulus aluminum\cite{grantaedupack}\\
			$\nu_{alu}$ & 0.33 $[-]$ & poisson's ratio  aluminum\cite{grantaedupack}\\
			$\rho_{alu}$ & 2700 $[kg/m^3]$ & density aluminum\cite{grantaedupack} \\
			$E_{alx}$ & 330 $[GPa]$ & young's modulus alumina\cite{grantaedupack}\\
			$\nu_{alx}$ & 0.22 $[-]$ & poisson's ratio  alumina\cite{grantaedupack}\\
			$\rho_{alx}$ & 3700 $[kg/m^3]$ & density alumina\cite{grantaedupack} \\
			$E_{fc}$ & 35 $[MPa]$ & young's modulus foam core\cite{trindade1999parametric}\\
			$\nu_{fc}$ & 0.3 $[-]$ & poisson's ratio  foam core\cite{grantaedupack}\\
			$\rho_{fc}$ & 32 $[kg/m^3]$ & density foam core\cite{grantaedupack}\\
			$E_{g}$ & 2.35 $[GPa]$ & young's modulus epoxy glue\cite{grantaedupack}\\
			$G_{g}$ & 0.84 $[GPa]$ & shear modulus epoxy glue\cite{grantaedupack}\\
			$\rho_{g}$ & 1410 $[kg/m^3]$ & density epoxy glue\cite{grantaedupack} \\
			$E_{fs}$ & 3.5 $[MPa]$ & estimated young's modulus foam spacers\\
			$\nu_{fs}$ & 0.3 $[-]$ & poisson's ratio  foam spacers\cite{grantaedupack}\\
			$\rho_{fs}$ & 427.37 $[kg/m^3]$ & foam spacer density, measured by volume \\
						&					& and mass of 1000x18x0.5 mm foam strip\\
			\hline 
		\end{tabular}
  }
	\end{center}
\end{table}
\section{Nyquist diagram OL-PPF control}
\label{Appendix:C_nyquist}

\begin{figure}[H]
	\begin{subfigure}[h]{0.49\textwidth}
		\centering
		\begin{tabular}{c}
			\includegraphics[height=4cm]{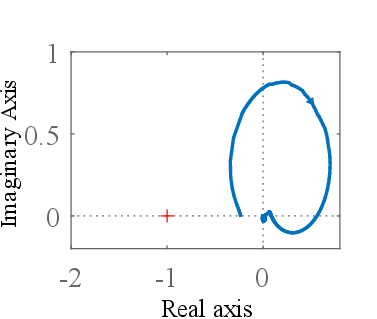}
		\end{tabular}
		\caption{Open-loop nyquist diagram of the system shown in Fig. \ref{fig:PPF_damped_b} (g = 0.3). Open-loop system is stable} 
		\label{fig:nyquist_a}
	\end{subfigure}
	\hfill
	\begin{subfigure}[h]{0.49\textwidth}
		\centering
		\begin{tabular}{c}
			\includegraphics[height=4cm]{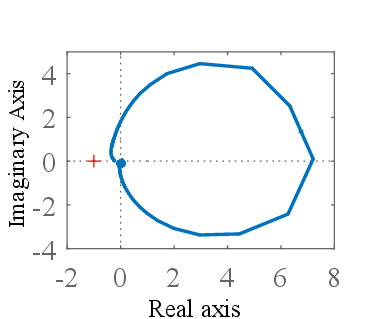}
		\end{tabular}
		\caption{Open-loop nyquist diagram of the system shown in Fig. \ref{fig:PPF_lowdamp_b} (g = 0.3). Open-loop system is stable }
		\label{fig:nyquist_b}
	\end{subfigure}
	\caption[example] 
	%>>>> use \label inside caption to get Fig. number with \ref{}
	{ \label{fig:nyquist} Nyquist diagram of the open-loop ($-K_{PPF}*H_{22}$).} 
\end{figure}

\end{document}